# Self-similar Magneto-electric Nanocircuit Technology for Probabilistic Inference Engines

Santosh Khasanvis[1*], Mingyu Li[1], Mostafizur Rahman[1], Mohammad Salehi Fashami[2], Ayan K. Biswas[2], Jayasimha Atulasimha[2], Supriyo Bandyopadhyay[2], and Csaba Andras Moritz[1+]

Department of ECE, University of Massachusetts Amherst[1], Amherst, MA, USA
Virginia Commonwealth University[2], Richmond, VA, USA
khasanvis@ecs.umass.edu[*], andras@ecs.umass.edu[+]

*Abstract*— **Probabilistic graphical models are powerful mathematical formalisms for machine learning and reasoning under uncertainty that are widely used for cognitive computing. However they cannot be employed efficiently for large problems (with variables in the order of 100K or larger) on conventional systems, due to inefficiencies resulting from layers of abstraction and separation of logic and memory in CMOS implementations. In this paper, we present a magneto-electric probabilistic technology framework for implementing probabilistic reasoning functions. The technology leverages Straintronic Magneto-Tunneling Junction (S-MTJ) devices in a novel mixed-signal circuit framework for direct computations on probabilities while enabling in-memory computations with persistence. Initial evaluations of the Bayesian likelihood estimation operation occurring during Bayesian Network inference indicate up to 127x lower area, 214x lower active power, and 70x lower latency compared to an equivalent 45nm CMOS Boolean implementation.**

*Keywords—Probabilistic graphical models; Bayesian networks; non-Boolean computing; mixed-signal; nanoscale; memory-in-computing.*

## I. INTRODUCTION

Most real-world computation problems e.g., graphics processing, network threat detection, medical diagnoses, speech recognition, data-mining, etc., require reasoning or decision-making in the presence of uncertainty: i.e., without the availability of complete information and/or well-characterized logic relationships. Probabilistic models (such as Bayesian Networks [1][2]) are a powerful formalism capable of reasoning under uncertainty and highly suitable for addressing such applications [3]-[5]. These models use probabilities as the basis of representing uncertainty in knowledge for a given domain, and require computations on probabilities for reasoning and machine learning. These tasks are performed for every variable involved in the domain and require (i) distributed storage of probabilities, and (ii) frequent arithmetic operations such as multiplication and addition of probability values. A key requirement for scalable hardware implementation of probabilistic graphical models is the efficient and parallel implementation of these probabilistic computations.

Conventional von Neumann architectures are not well suited because they (i) would require emulation of an inherently non-deterministic, non-logical computing model on a deterministic Boolean logic framework, (ii) incorporate a limited number of arithmetic units (due to high complexity of implementing resource-intensive operations such as multiplication with Boolean logic) which leads to serialized execution (even with multi-core processors) for models with large number of variables, (iii) use a rigid separation between logic and memory, as opposed to supporting distributed local storage and processing capabilities, and (iv) use a radix-based representation of data which is inefficient for representing probabilistic information and has no inherent fault-resilience.

We propose a new non-Boolean multi-domain mixed-signal circuit framework for probabilistic computation at nanoscale, called Probability Arithmetic Composers. This is applicable in reasoning and machine learning frameworks that use probabilistic graphical models for knowledge representation, such as Bayesian Networks (BNs). The main contributions include: (i) an unconventional multi-valued spatial probabilistic information representation supporting graceful degradation, (ii) a new mixed-signal Probability Arithmetic Composer circuit framework to implement arithmetic operations on probabilities while supporting memory-in-computation, where elementary arithmetic functions themselves are the building blocks instead of logic functions, and (iii) evaluation of proposed approach vs. CMOS implementation of likelihood estimation operation for Bayesian Network inference as an example. The Probability Arithmetic Composer paradigm utilizes voltage-controlled Straintronic MTJ (S-MTJ) devices, where the applied input voltage results in a strain-induced magnetization reorientation in the S-MTJ free layer, which can be made persistent for non-volatility. This magnetization reorientation changes the S-MTJ resistance that can be measured with tunneling current through the device, generated by a reference voltage. Thus the S-MTJ provides a mechanism for efficient compression of redundant information in magnetic domain (resistance) into a compact form (current/voltage) for computation. While we use binary S-MTJs as an example in this paper, S-MTJs may be designed with multiple magnetization states to enable new multi-valued redundant representation of information, which can easily be converted into probability values and supports graceful degradation in the presence of errors vs. conventional radix representations. Other devices that exhibit such multi-domain interactions with non-volatility may also be used.

The rest of the paper is organized as follows. The underlying technology using voltage controlled S-MTJ device and spatial probabilistic data representation are described in Sections II and III respectively. Section IV presents an overview of the new mixed-signal Probability Arithmetic

This material is based upon work supported by the National Science Foundation grant 1407906 at UMass Amherst, and National Science Foundation grants ECCS-1124714, CCF-1216614 and CCF-1253370 at VCU.

Composer circuit framework for computation on probabilities. Section V presents details on circuits for elementary arithmetic functions on probabilities as building blocks for the Probability Arithmetic Composer framework. Section VI presents an overview of Bayesian Networks as an application example. Section VII describes the evaluation methodology and comparison with conventional CMOS implementations, followed by conclusion in Section VIII.

## II. TECHNOLOGY OVERVIEW: VOLTAGE CONTROLLED STRAINTRONIC MTJ DEVICE

The concept of straintronics, where the bistable magnetization of a shape anisotropic multiferroic nanomagnet is switched with electrically generated mechanical strain, is attractive due to its extreme low energy of switching. A straintronic MTJ (S-MTJ) device is shown in Figure 1a. It consists of three layers - a "hard" ferromagnetic layer with a fixed magnetization orientation, an ultrathin spacer layer, and a "soft" ferromagnetic layer with variable magnetization orientation. The three layered stack is fabricated on a thin piezoelectric film grown on an $n^+$-Si substrate.

Because of dipole coupling between the hard and soft layers, they tend to have mutually anti-parallel magnetizations (see Figure 1a) and in that configuration, the resistance of the S-MTJ measured between the two ferromagnetic layers is *high*. Application of an input voltage ($V_{in}$) at the two (shorted) contact pads generates a biaxial strain in the piezoelectric layer underneath the soft magnet (compression along the major axis of the elliptical soft magnet and tension along the minor axis) [7][8], which rotates the magnetization of the soft magnet by an angle $\Theta$ via the Villari effect, if the soft layer is magnetostrictive and has positive magnetostriction. This reduces the angular separation between the magnetization orientations of the hard and soft layers, which in turn reduces the resistance of the S-MTJ. If the input voltage is withdrawn, the stress in the soft magnetic layer relaxes and hence its magnetization will tend to return to its original orientation because of dipole coupling with the hard magnetic layer. In this case, the operation is volatile. The resistance ratio between the high- and low-resistance states as a function of applied voltage $v$ is roughly given by [9],

$$r(v) = \frac{R_{ON}}{R_{OFF}} = \frac{R(v=V_{ON})}{R(v=0)} = \frac{1-\eta_1\eta_2}{1-\eta_1\eta_2 \cdot \cos[\Theta(V_{ON})]}, \quad (1)$$

where $\Theta(V_{ON})$ is the angle by which the magnetization of the soft layer rotates under stress generated by input voltage $V_{ON}$, assuming it starts from being exactly anti-parallel to the hard layer initially, and $\eta_1$, $\eta_2$ are the spin-injection/filtering efficiencies at the interfaces between the two ferromagnets and the spacer layer. At room temperature, these quantities are roughly 70% [10]. The maximum value of $\Theta$ is $90^0$ unless the input voltage pulse is timed in a certain way to allow reorientation by $180^0$ [11].

The magnetization rotation can be made persistent through a scheme shown in Figure 1b, resulting in non-volatile operation. The electrodes $A - A'$ are shorted to form one input terminal, and $C - C'$ are shorted to form the second terminal. When a voltage is applied between these terminals and the $n^+$-substrate, electric fields are generated underneath the pads, producing a highly localized strain field in the piezoelectric film [7][8]. This results in biaxial strain (compression/tension along the line joining the electrodes and tension/compression along the perpendicular direction) since the distance between the electrode pairs is approximately equal to the PZT film thickness. This strain will then be elastically transferred to the soft layer of the S-MTJ stack despite any substrate clamping. The scheme requires a small in-plane external magnetic field (***B***) along the minor axis of the soft magnet which brings the two stable magnetization states out of the soft magnet's major axis (easy axis) and aligns them along two in-plane directions that lie between the major and minor axes with an angular separation of ~$132^0$. These two stable orientations ($\Psi_1$ and $\Psi_0$) of magnetization represent the *low* and *high* resistance states, respectively. The magnetization of the hard magnetic layer is parallel to $\Psi_1$, which is why the low resistance state is visited when the magnetization of the soft magnetic layer is along $\Psi_1$. Since Terfenol-D has a positive magnetostriction coefficient, compressive stress along the line joining the electrodes *A–A'* will stabilize the magnetization at $\Psi_0$, while a compressive stress along *C–C'* electrodes will switch the magnetization back to $\Psi_1$ [15]. These magnetization orientations are stable, i.e. if the magnetization is left in either state it remains there in perpetuity even after power is switched off, which makes the device non-volatile. The change in resistance of the S-MTJ is read by using a reference voltage, which generates an output current. Thus, conversion between voltage, magnetic and current domains is achieved.

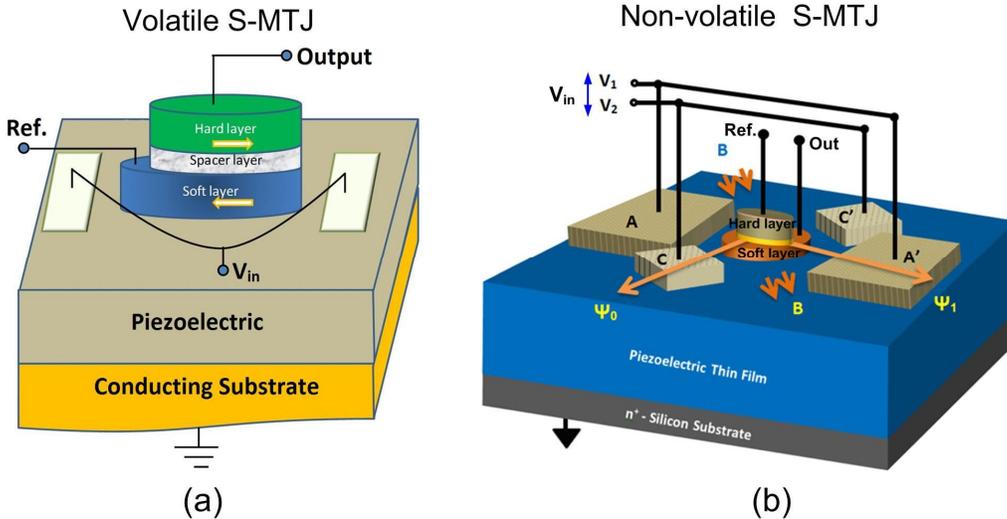

Figure 1. (a) Volatile S-MTJ device configuration: Voltage input induces strain in soft-layer layer adjusting magnetization orientation; a reference terminal (Ref.) is used for resistance readout; and (b) Non-volatile S-MTJ device: The MTJ stack is placed in between two pairs of electrode pads such that the line joining each electrodes subtends an angle of $15^0$ and $165^0$ respectively with the major axis of soft magnetic layer. A magnetic field ***B*** is applied along the minor axis of the soft magnetic layer. Voltage input persistently changes magnetization orientation through strain.

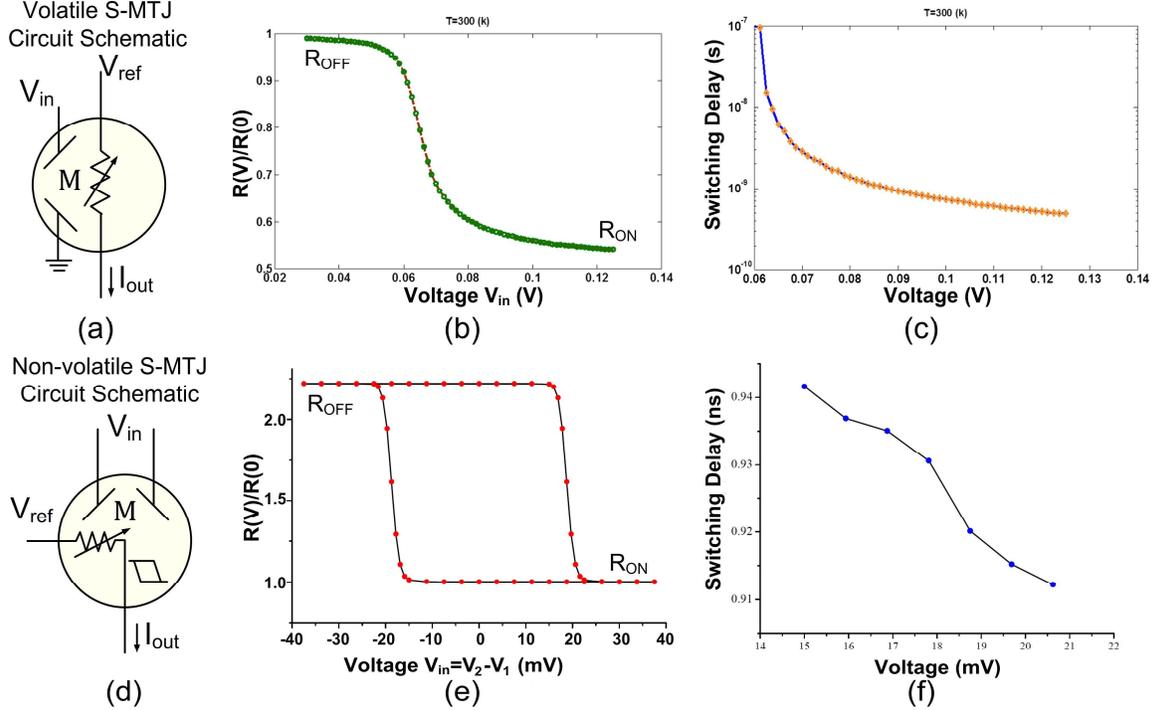

Figure 2. (a) Volatile S-MTJ circuit schematic; (b) Simulated DC transfer characteristics for volatile S-MTJ showing resistance ratio *r(v)*, as function of input voltage $V_{in}$; (c) Simulated switching delay characteristics for volatile S-MTJ; (d) Non-volatile S-MTJ circuit schematic; (e) Simulated DC transfer characteristics for non-volatile S-MTJ showing resistance ratio *r(v)*, as function of input voltage $V_{in}$. Hysteresis indicates persistence in resistance state; and (c) Simulated switching delay characteristics for non-volatile S-MTJ.

The transfer characteristics of the S-MTJ devices (Figures 2b-c and Figures 2e-f) are extracted from stochastic Landau-Lifshitz-Gilbert (LLG) simulations, described in refs. [12]-[16]. For the volatile S-MTJ transfer characteristics, we used a soft layer made of Terfenol-D with dimensions 120nm x 105nm x 6.5nm, and 110nm x 90nm x 9 nm for non-volatile S-MTJ. The piezoelectric layer was assumed to be lead-zirconate-titanate (PZT) of thickness 100nm. The effect of room-temperature thermal noise was taken into account [12]-[16] and the characteristics presented are thermally averaged characteristics. Furthermore, although the strain generated in the magnet is biaxial, we approximated it with uniaxial strain (which overestimates the voltage needed to generate a given strain). This is somewhat compensated by the fact that we assume 100% strain transfer from the piezoelectric film to the magnetostrictive layer, leading to an underestimation of the voltage needed to generate a given strain. Every data-point in Figures 2b,e is generated by averaging 10,000 simulations. The LLG simulations also yield the switching time needed for $\Theta(v)$ to stabilize to its final value after input voltage is abruptly switched on, shown in Figures 2c,f.

### III. MULTI-VALUED PROBABILISTIC INFORMATION REPRESENTATION

In the proposed framework, information is represented in the magnetic domain (magnetization vector orientation of the S-MTJ free-layer, and thus the resistance) as a non-Boolean probabilistic vector of '*n*' spatially distributed digits $(p_1, p_2, \ldots, p_n)$. As opposed to conventional number systems (e.g. binary, HEX etc.), in this representation all digits carry equal weight irrespective of position, which implies inherent redundancy and better error resilience through graceful degradation. Each digit $p_i$ can take any one of '*k*' values, where *k* is the number of distinct magnetization states for an S-MTJ (e.g., for S-MTJs with 4-states, *k*=4 and any digit $p_i$= {0,1,2,3}). The value of the probability ***P*** represented by this vector is given by the following formula:

$$\boldsymbol{P} = \frac{\sum_{i=1}^{n} p_i}{n(k-1)}. \tag{2}$$

Here, each digit $p_i \in \{0,1,\ldots,k\text{-}1\}$. Each probability digit is physically represented in a persistent manner using the non-volatile S-MTJ resistance states, determined by relative magnetization orientation of the magnetic layers. For example, for binary S-MTJ devices the probability digit 0 is represented using *high* resistance and digit 1 is represented with *low* resistance. Since these resistance states are programmed using input voltages, there is a corresponding digital voltage equivalent for probability representation. The data is read-out in analog electrical domain with discrete current/voltage values as explained later in this paper (see Figure 3 for equivalent data representations in multiple domains for the case of binary S-MTJs as an example). The *resolution* of data representation and computation is defined as the minimum non-zero probability value that can be represented in this format, determined by the number of digits $n$ and the number of states of each digit $k$.

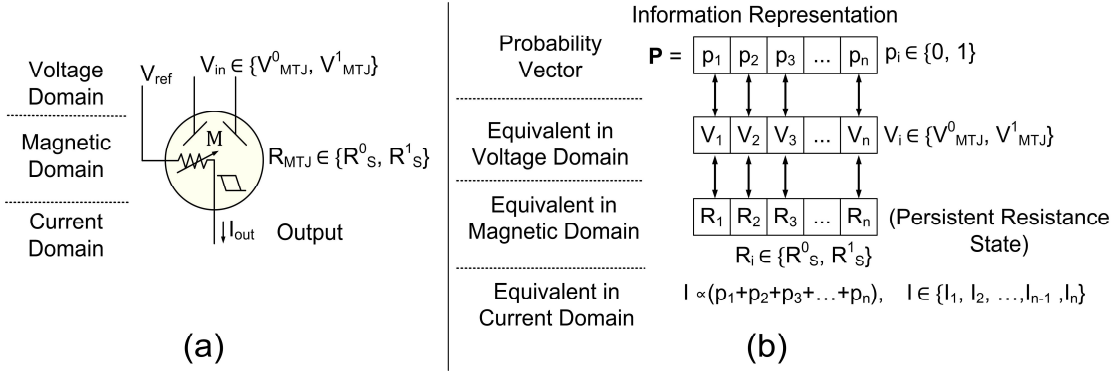

Figure 3. (a) Non-volatile S-MTJ device schematic showing multi-domain representation: $V_{in}$ is the input voltage between the two input terminals for switching magnetization, $V_{ref}$ is used during readout; and (b) Spatial probabilistic information representation for S-MTJ with 2 states, and its equivalent in resistance, voltage and current domains.

### A. Fault Resilience (Supporting Graceful Degradation)

Information representation and computation in our approach is inherently fault resilient (with graceful degradation) in both electrical and magnetic domains. Consider two possible single-fault scenarios: (i) input voltage at any position is shifted by a single level, and (ii) a magnetization vector in a S-MTJ is offset to a neighboring state of the 'intended' value. Given that the representation is redundant with all digits carrying equal weight, either fault would cause the overall value to be erroneous by $1/[n(k-1)]$, i.e., the resolution of the computation. This is in direct contrast to conventional $n$-bit radix-based representations where a single fault can cause up to a $2^{n-1}$ error in the value being stored/computed. Our approach thus supports graceful degradation, which is linear with increasing number of faults. Furthermore, the number of digits used ($n$) can be adjusted depending on the precision and fault-resilience required by the application.

### IV. PROBABILITY ARITHMETIC COMPOSER FRAMEWORK

We propose an unconventional mixed-signal *Probability Arithmetic Composer* circuit framework for probabilistic computation, using emerging nanoscale devices exhibiting multi-domain interactions (S-MTJ devices) and multi-valued probabilistic information representation. Here, arithmetic functions themselves are the basic building blocks, rather than relying on Boolean logic. A Probability Arithmetic Composer performs arithmetic operations on probabilities that are in spatial probabilistic representation encoded in multiple resistance states (magnetic domain). The result of the operation is in multi-valued discrete electrical (analog current/voltage) domain. A *Decomposer* circuit is used to convert back to a redundant spatial representation for cascading successive Arithmetic Composers and/or interfacing with CMOS.

A Probability Arithmetic Composer can be recursively defined as a hierarchical instantiation of other Arithmetic Composer functions until Elementary Arithmetic Composer functions with S-MTJs are reached, as shown in Figure 4. Thus, a Probability Arithmetic Composer ($f^n$) consisting of

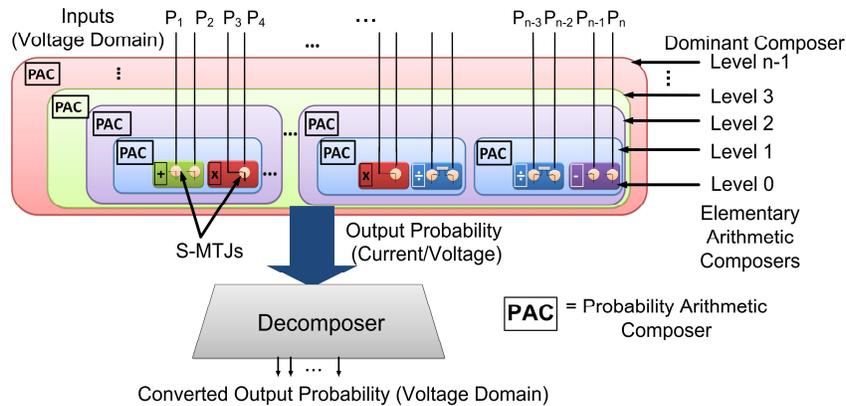

Figure 4. Probability Arithmetic Composer Circuit Framework: Hierarchical representation of Probability Arithmetic Composers showing nested levels of self-similar Composers, with top-most level ($n$-1) being Dominator Composer innermost (level-0) being Elementary Arithmetic Composers.

'*n*' levels of operations to be performed can be recursively expressed as:

$$\text{for } n > 1, \ f^n = f^{n-1}(f_1^{n-2}, f_2^{n-2}, f_3^{n-2}, \ldots, f_j^{n-2}) \quad (3)$$

$$\text{for } n = 1, \quad f^1 = f^0(primary\ inputs). \quad (4)$$

Here $f^0$ is Elementary Arithmetic Composer acting on primary inputs. The top-level operation to be performed ($f^{n-1}$) is called the Dominator Composer since it determines the overall Arithmetic Composer circuit topology, where each component is either another Arithmetic Composer or an Elementary Arithmetic Composer. This approach is easily scalable since any Arithmetic Composer can be hierarchically built by plugging Arithmetic Composer nodes in a Dominator Arithmetic Composer without changing the circuit style, leading to *self-similar* fractal-like circuits.

For example, a function F = $(P_a.P_b)+(P_c.P_d)$ can be hierarchically represented as F = $f^2 = f^1(f_1^0, f_2^0)$ = SUM[MUL($P_a$, $P_b$), MUL($P_c$, $P_d$)]. Here $n=2$ since there are two levels of operations to be performed ($f^1 = SUM$ and $f^0 = MUL$). While S-MTJs are used in this work, the framework is generic and any other device exhibiting multi-domain interactions and non-volatility may be used as well.

*A. Probability Composer Circuit*

We use a *Probability Composer* circuit to convert the spatial probability representation in magnetic domain (resistance) to the electrical domain for computation. The output can be either in analog current or voltage domains, and is readout by using a reference voltage. For an *n*-digit probability vector, a Probability Composer uses $n$ S-MTJ devices each having $k$ states (Figure 5a). The output of the circuit is proportional to the sum of all inputs and has [$n(k-1)$ + 1] distinct resistance states. Thus the output resolution is $1/[n(k-1)]$. As an example, the output resistance states for a Probability Composer using 10 binary S-MTJs (i.e. $n=10$, $k=2$) is shown in Figure 5b for a resolution of 0.1. We use an inverse-linear relationship between S-MTJ resistance ($r_i$) and the probability digit ($p_i$) as follows.

$$r_i = \frac{\beta}{p_i + \varepsilon}. \quad (5)$$

Here, $\beta$ and $\varepsilon$ are constants chosen such that the above relationship holds. For binary devices with two resistance states ($r_i=R_{OFF}$ corresponding to $p_i=0$ and $r_i=R_{ON}$ corresponding to $p_i=1$), by substituting the corresponding $r_i$ and $p_i$ values we get

$$\varepsilon = \frac{1}{\left(\frac{R_{OFF}}{R_{ON}}-1\right)}; \ \beta = \varepsilon \cdot R_{OFF} = \frac{R_{OFF}}{\left(\frac{R_{OFF}}{R_{ON}}-1\right)}. \quad (6)$$

Alternative representations may also be used where the resistance is linear with respect to the probability digit. Such alternatives will require changes to the circuit implementations as well. The effective resistance of an *n*-digit Probability Composer ($R_{PC}$) using binary S-MTJs has $n+1$ discrete states, given by the following expression. Here, the inverse of $R_{PC}$ is proportional to the sum of probability digits.

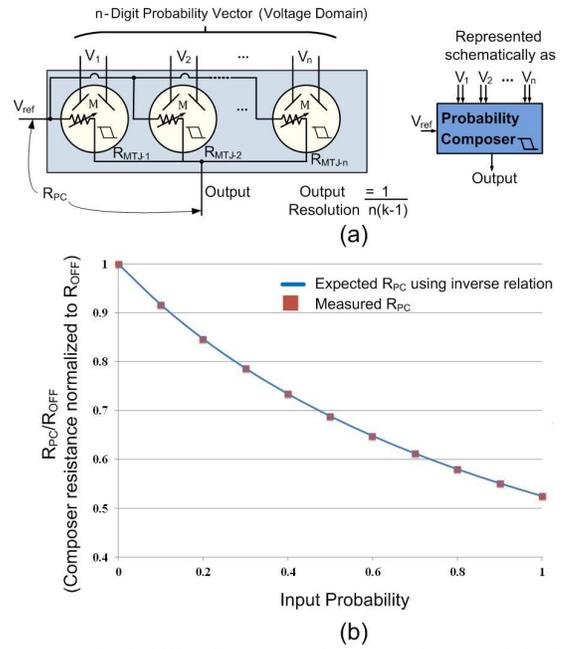

Figure 5. (a) Probability Composer circuit topology; and (b) The effective resistance for corresponding encoded probability value for binary S-MTJ as an example (represented using probability digits and stored in each S-MTJ resistance state). Resistance is normalized to its OFF state resistance.

$$\frac{1}{R_{PC}} = \sum_{i=1}^{n}\frac{1}{r_i} = \sum_{i=1}^{n}\frac{(p_i+\varepsilon)}{\beta} = \frac{1}{\beta}\left[\sum_{i=1}^{n}p_i\right] + \left\{\frac{n\varepsilon}{\beta}\right\}. \quad (7)$$

When using a load resistance $R_L$ much smaller than the S-MTJ resistance connected between the output terminal of the Probability Composer and ground, the output current flowing through this load resistor is given by:

$$\begin{aligned}I_{out} &= \frac{V_{REF}}{(R_{PC}+R_L)} \approx \frac{V_{REF}}{R_{PC}}, R_L \ll R_{PC} \\ &= \frac{V_{REF}}{\beta}\left[\sum_{i=1}^{n}p_i\right] + \left\{\frac{n\varepsilon V_{REF}}{\beta}\right\}.\end{aligned} \quad (8)$$

The term in {.} represents the additional current that needs to be corrected for output linearity. This can be done with a Correction Circuit (see Figure 6a), such that the output current is given by:

$$\begin{aligned}I_{out} &\approx \frac{V_{REF}}{\beta}\left[\sum_{i=1}^{n}p_i\right] + \left\{\frac{n\varepsilon V_{REF}}{\beta}\right\} + \frac{V_{ADJ}}{R_{ADJ}} \\ &= \frac{V_{REF}}{\beta}\left[\sum_{i=1}^{n}p_i\right] = \frac{nV_{REF}P}{\beta}.\end{aligned} \quad (9)$$

Here, $V_{ADJ} = -V_{REF}$, $R_{ADJ} = \beta/(n.\varepsilon)$ and **P** is the probability value represented by the digital probability vector as defined in equation (2). Thus for every probability value there is a corresponding current domain output. However, we are interested in a voltage output since S-MTJs are voltage controlled. The current domain signal can be converted to analog voltage domain by using the resultant voltage across the load resistance. However, since the value of $R_L$ has to be necessarily low relative to S-MTJ resistance for the approximation in equations (8)-(9), the range of output voltages using this scheme is too low to be useful without significant amplification. If the output voltage non-linearity can be tolerated at read-out (through the use of *Decomposer* circuits explained next), then the analog voltage output with a larger range can be obtained by simply eliminating the load resistance $R_L$ (see Figure 6c). The output voltage is given by the following expression:

$$V_{out} = V_{REF} \cdot \left[ \frac{\frac{1}{R_{PC}} - \frac{1}{R_{ADJ}}}{\frac{1}{R_{PC}} + \frac{1}{R_{ADJ}}} \right] = V_{REF} \cdot \left[ \frac{\sum_{i=1}^{n} p_i}{\sum_{i=1}^{n} p_i + 2n\varepsilon} \right] \quad (10)$$
$$= V_{REF} \cdot \left[ \frac{P}{P + 2\varepsilon} \right].$$

Here $P$ is the probability value represented by the digital probability vector, defined in equation (2). This topology results in a non-linearity in the output; for probability values close to 0 the output voltage is proportional to sum of individual probability digits, but degrades for probability values close to 1. As long as different output levels can be differentiated, the above topology may be used. This represents a trade-off between using subthreshold CMOS analog support circuits for amplifying the low output voltage range exhibiting linearity as in the case with current-mode readout, vs. tolerating non-linearity in output for wider voltage range with a potentially simpler circuit implementation for voltage-mode readout. This circuit can now be considered as an element with higher resolution than a single S-MTJ and can be used for building high-resolution circuits for probability arithmetic.

*B. Decomposer Circuit*

We need an approach to convert the analog voltage output at a Composer circuit back to a digital probability vector representation. To achieve this we design a *Decomposer* circuit with volatile S-MTJs as follows (see Figure 7). The Decomposer has the following requirements: (i) For converting analog input voltage to an *n*-digit probability vector, it requires *n Decomposer Elements*; each Decomposer Element is designed to trigger at a different input voltage value, i.e. they have different threshold voltages. (ii) When triggered, each Decomposer Element needs to generate a pair of differential output voltage signals, so as to switch a non-volatile S-MTJ in the successive stage.

Drawing inspiration from flash analog-to-digital converters, we use a resistive ladder to setup varying threshold voltages for each decomposer element (see Figure 7b). When uniform resistances are used in the ladder, it responds to a linear change in input voltage. Input non-linearity can be accommodated by using non-uniform resistances in the ladder. Alternatively, the S-MTJ device may be designed to have varying thresholds by changing the device parameters (such as PZT thickness, etc.). Here, the volatile S-MTJs in each decomposer element (see Figure 7a) act as voltage comparators; if the input voltage is above the control voltage (setup using the resistance ladder) the S-MTJ switches its resistance; else it remains in its previous state. To generate differential voltage output when triggered, each decomposer element consists of two branches; one with S-MTJ in pull-up and the other with S-MTJ in pull-down. The possible states of the S-MTJs and the corresponding output voltages are shown in Table I for this configuration.

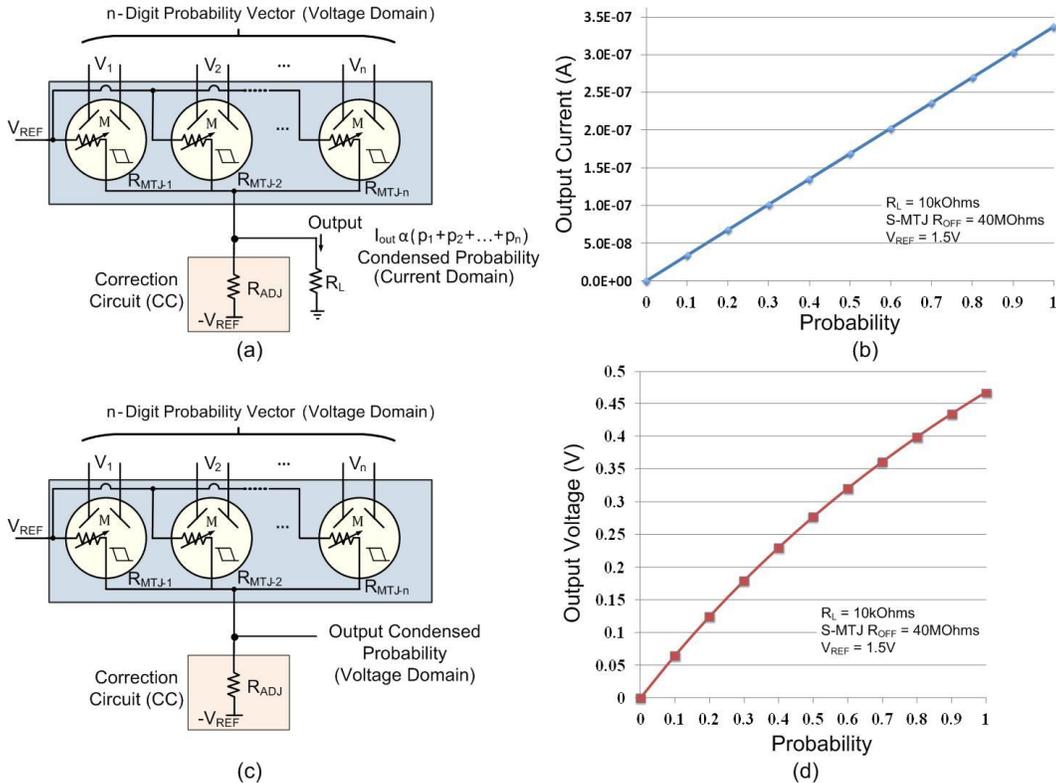

Figure 6. Read-out schemes for Probability Composer circuit. (a) Current-mode read-out with corresponding output values shown in (b); and (c) Voltage-mode read-out with corresponding values shown in (d).

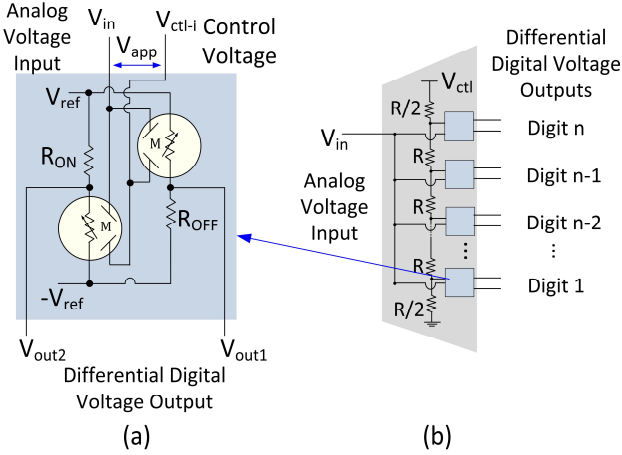

(a) (b)

Figure 7. Decomposer Circuit Design: (a) Decomposer Element used to generate differential digital voltages based on analog input voltage for a given threshold voltage; and (b) Full Decomposer circuit consisting of *n* Decomposer Elements to convert analog voltage signal to *n*-digit probability vector using discrete voltage representation. Here, $V_{ctl-i}$ controls the threshold voltage for the *i*-th element and is determined by the resistance ladder network.

## V. ELEMENTARY ARITHMETIC COMPOSERS

Elementary Arithmetic Composers are the circuits at the lowest level of the recursive Probability Arithmetic Composer definition that perform atomic arithmetic operations on probabilities. In these circuit implementations, we leverage underlying physical laws for computation rather than using abstraction with Boolean logic. Addition on two probabilities (represented using two Probability Composers) is easily performed by using a parallel topology of the two Probability Composers. Since the inverse of the resistance of each Probability Composer is proportional to the encoded probability value, the inverse of the total resistance is proportional to the sum of the two input probabilities. By applying a common reference voltage and using correction

TABLE I. DECOMPOSER ELEMENT OPERATION

| Operating Condition $V_{app} =$ ($V_{in} - V_{ctl}$) | S-MTJ Resistance | Output1 ($V_{out1}$) | Output2 ($V_{out2}$) | Probability Digit |
|---|---|---|---|---|
| $V_{app} < V_{th}$ | $R_{OFF}$ | 0 | $V_{REF}/3$ | 0 |
| $V_{app} \geq V_{th}$ | $R_{ON}$ | $V_{REF}/3$ | 0 | 1 |

Note: Here, $V_{in}$ is the analog input voltage applied to the decomposer circuit, $V_{app}$ is the applied voltage difference across the inputs of a Decomposer Element, and $V_{th}$ is the threshold voltage of switching for a decomposer element.

circuits as before to overcome the limited $R_{OFF}/R_{ON}$, the output can be read either in current or voltage domain as discussed below.

$$I_{out} = \frac{V_{REF}}{\left(\frac{R_{PC-A} \cdot R_{PC-B}}{R_{PC-A} + R_{PC-B}} + R_L\right)} \approx V_{REF}\left(\frac{1}{R_{PC-A}} + \frac{1}{R_{PC-B}}\right)$$

$$= \frac{V_{REF}}{\beta}\left[\left(\sum_{i=1}^{n} p_i\right)_A + \left(\sum_{i=1}^{n} p_i\right)_B\right] + \left\{\frac{2n\varepsilon V_{REF}}{\beta}\right\}$$

$$= \frac{nV_{REF}}{\beta}[\boldsymbol{P_A} + \boldsymbol{P_B}] + \left\{\frac{2n\varepsilon V_{REF}}{\beta}\right\}; \; R_L \ll R_{PC}. \quad (11)$$

Alternatively, the voltage mode-readout for a single Probability Composer can be extended for two inputs by using each additional Probability Composers in parallel, as shown in Figure 8a. This scheme however exhibits non-linearity in output voltage (Figure 8b). Extending this parallel topology for all Probability Composers involved easily accommodates scaling to a higher number of inputs.

$$V_{out} = V_{REF} \cdot \frac{\left[\frac{1}{R_{PC-A}} + \frac{1}{R_{PC-B}} - \frac{2}{R_{ADJ}}\right]}{\left[\frac{1}{R_{PC-A}} + \frac{1}{R_{PC-B}} + \frac{2}{R_{ADJ}}\right]} \quad (12)$$

$$= V_{REF} \cdot \left[\frac{\boldsymbol{P_A} + \boldsymbol{P_B}}{\boldsymbol{P_A} + \boldsymbol{P_B} + 4\varepsilon}\right]$$

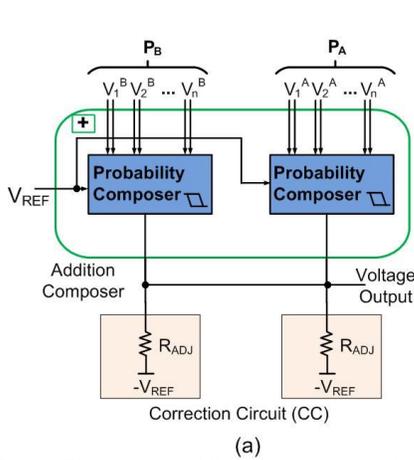

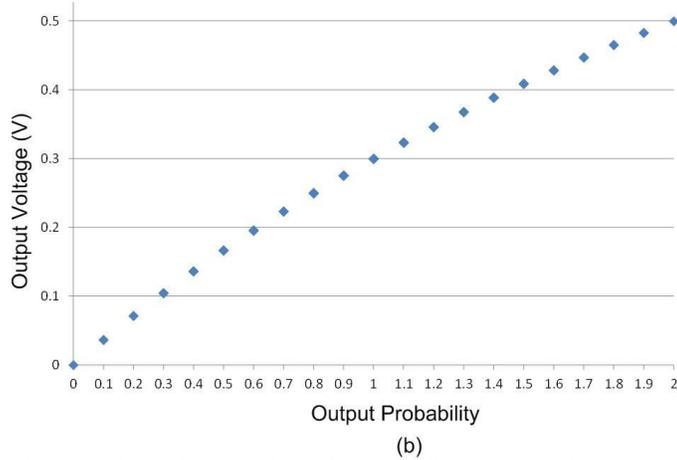

(a) (b)

Figure 8. (a) Elementary addition composer using voltage mode read-out; and (b) Corresponding output voltage vs. probability characteristics as calculated by eq. (12), and validated using HSPICE simulations for all possible input combinations.

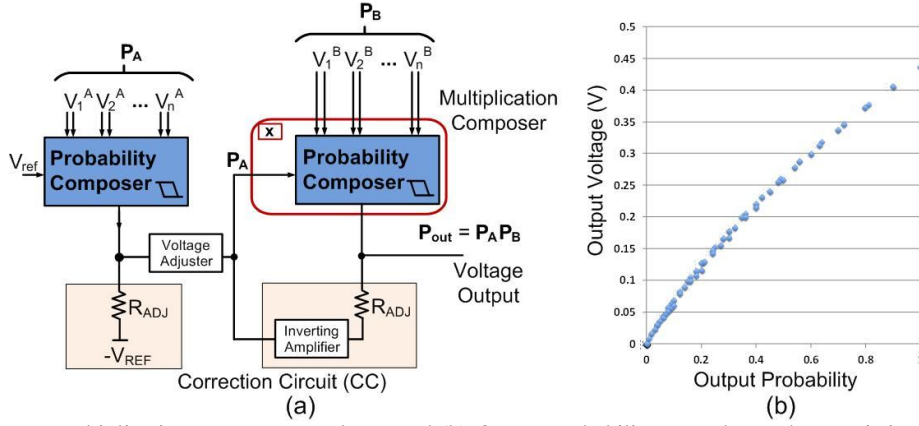

Figure 9. (a) Elementary multiplication composer topology; and (b) Output probability vs. voltage characteristics in continuous analog domain validated using HSPICE simulations for all possible input combinations.

To implement multiplication, we leverage Ohm's law, $V = I.R$. Rewriting the relationship as $I = V.(1/R)$, we use one set of inputs for multiplication in voltage domain ($V$) and the other in the resistance ($1/R$), since the inverse of the Probability Composer's resistance is proportional to the stored probability value. For the first input in voltage domain, we use another Probability Composer storing the input probability value and voltage-mode readout. The Multiplication Composer topology is shown in Figure 9a. Thus the resulting output current represents multiplication of two probabilities. In the following equations, $g$ represents the gain of voltage amplifier (e.g. using CMOS analog support circuits such as op-amps) used to amplify the output of first Probability Composer.

$$I_{out} \approx \frac{n}{\beta}[V_A \cdot \boldsymbol{P_B}] = \frac{n}{\beta}.g.V_{REF}.\left(\frac{P_A P_B}{P_A + 2\varepsilon}\right); R_L \ll R_{PC}. \quad (13)$$

Eliminating the load resistance $R_L$ and simply using the output node voltage achieves voltage-mode readout with a higher output range. The output voltage is governed by the following relation:

$$V_{out} = g.V_{REF}.\left[\frac{\boldsymbol{P_A} \cdot \boldsymbol{P_B}}{(\boldsymbol{P_A} + 2\varepsilon)(\boldsymbol{P_A} + 2\varepsilon)}\right] \quad (14)$$

A more complex arithmetic operation such as sum-of-products can be composed using these Elementary Addition and Multiplication Composers. We illustrate an example to compose an operation of the form $(P_A.P_B)+(P_C.P_D)$. One way to implement it is to use Elementary Addition and Multiplication Composers and connect them serially. However, the Probability Arithmetic Composer framework allows us to implement it efficiently for parallel computation by hierarchically composing an Add-Multiply composer as follows. Each product term implemented with an elementary Multiplication Composer is arranged in a topology of the Addition Composer (see Figure 10a). Thus the dominator Composer structure is that of the adder, which uses elementary Multiplication composers as the basic building blocks. This topology realizes the add-multiply operation in a single step (simulated output characteristics shown in Figure 10b).

## VI. NANOSCALE COGNITIVE REASONING WITH BAYESIAN NETWORKS

A Bayesian Network (BN) is a probabilistic reasoning model [1][2] representing knowledge of an uncertain domain, whose structure (e.g., a tree) captures qualitative relationships between variables. This is attractive because it is a consistent

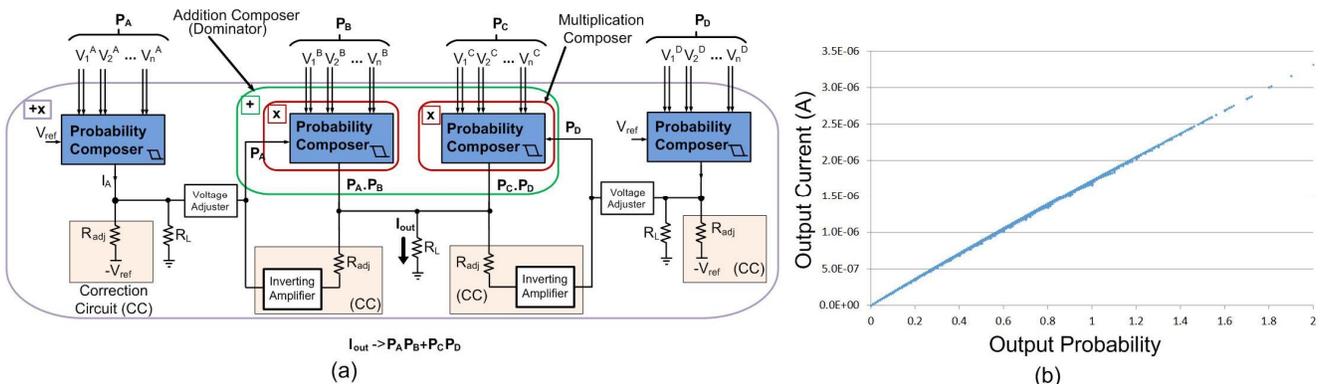

Figure 10. (a) Add-Multiply Composer for calculating sum-of-products on input probabilities. The output is in analog current-domain, and corresponds to the function, $P_A.P_B+P_C.P_D$. The voltage adjusters are used to amplify the voltage from first Probability Composer stage, which is then used as input voltage for read-out at the second stage. These adjusters and other support circuits such as the inverting amplifiers can be implemented using CMOS analog circuits (e.g. op-amps); and (b) Output characteristics showing probability output for all possible input combinations and the corresponding output current value, which are obtained using HSPICE simulations.

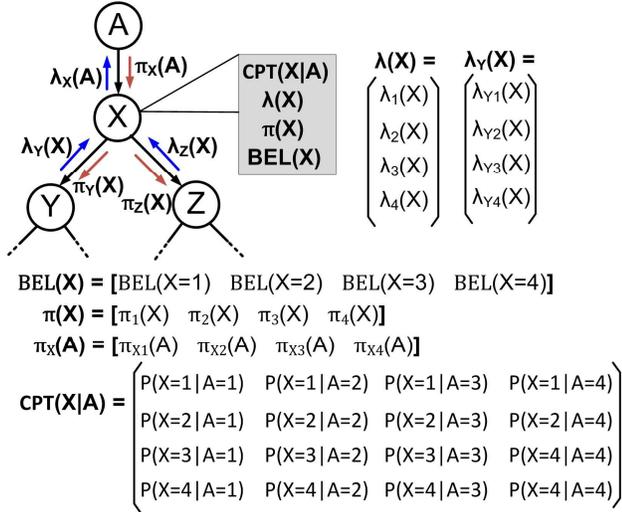

BEL(X) = [BEL(X=1)  BEL(X=2)  BEL(X=3)  BEL(X=4)]
π(X) = [π₁(X)  π₂(X)  π₃(X)  π₄(X)]
π_X(A) = [π_{X1}(A)  π_{X2}(A)  π_{X3}(A)  π_{X4}(A)]

$$CPT(X|A) = \begin{pmatrix} P(X=1|A=1) & P(X=1|A=2) & P(X=1|A=3) & P(X=1|A=4) \\ P(X=2|A=1) & P(X=2|A=2) & P(X=2|A=3) & P(X=2|A=4) \\ P(X=3|A=1) & P(X=3|A=2) & P(X=3|A=3) & P(X=4|A=4) \\ P(X=4|A=1) & P(X=4|A=2) & P(X=4|A=3) & P(X=4|A=4) \end{pmatrix}$$

Figure 11. Part of a BN with showing node X with parent A and child nodes Y, Z. All nodes have four states in this example. Each node maintains likelihood vector (**λ**), prior vector (**π**), belief vector (**BEL**), and conditional probability table (**CPT**). The CPT information and messages from child/parent nodes based on observed evidence are used to calculate **λ, π,** and **BEL** vectors during Bayesian inference.

and complete representation, in addition to being modular and compact [1][2]. A typical BN is a directed acyclic graph, with individual nodes representing knowledge about variables in a system. Dependencies between the variables are represented as directed links between the nodes. A node is a parent of a child if there is a directed link from former to the latter. A node without parents is called a root node, while a node without children is called a leaf node. Each node can have several states for its corresponding variable, and a conditional probability table (CPT) stores conditional probabilities that quantify the strength of its dependencies with its parents. A part of a typical BN is shown in Figure 11 focusing on a node *X* with one parent node *A* and two child nodes, *Y* and *Z*.

When constructing a BN for a specific application, hypotheses can be expressed as BN variables and a unique probability is assigned to each hypothesis initially (either based on prior knowledge of the domain or learned from data). Inference in the BN requires computation of belief, i.e. the probability of a hypothesis based on current events observed (state of evidence nodes) and corresponding conditional probability distributions, and is performed via message propagation (likelihoods and priors [1][2]) through the network using Pearl's Belief Propagation algorithm [1]. The key operations in a BN during inference are likelihood/prior estimation to generate these messages and belief update, which involve probability arithmetic.

Belief update refers to estimating the probability that a node is in a particular state based on evidence. For example, assuming every node in Figure 11 has four possible states, the belief update is performed as per equation (15). Here, variables in bold typeface represent matrices, the ⊗ operator represents matrix multiplication, and asterisk (*) represents element-wise multiplication. Likelihoods (**λ**) are represented as column vectors and priors (**π**) as row vectors (see Figure 11).

$$BEL(X) = \alpha \pi(X) * \lambda(X). \quad (15)$$

Here $\alpha$ is a normalization constant to ensure that the result is a probability. Likelihood estimation (**λ(X)**) and prior estimation (**π(X)**) for a node *X* are performed based on messages from its child nodes (**λ_Y(X), λ_Z(X)**) and parent node (**π_X(A)**) as follows:

$$\lambda(X) = \lambda_Y(X) * \lambda_Z(X) \quad (16)$$
$$\pi(X) = \pi_X(A) \otimes CPT(X|A). \quad (17)$$

Equation (17) has a sum-of-products form when expanded. Finally support messages to parent node (**λ_X(A)**) and child nodes (**π_Y(X), π_Z(X)**) are calculated as follows:

$$\lambda_X(A) = CPT(X|A) \otimes \lambda(X), \quad (18)$$
$$\pi_Y(X) = \alpha \pi(X) * \lambda_Z(X), and \quad (19)$$
$$\pi_Z(X) = \alpha \pi(X) * \lambda_Y(X).$$

All the above use multiplication and sum-of-product operations on probabilities, which can be implemented using Composers shown earlier. The likelihood estimation operation represented by eq. (16) can be implemented using multiplication composers, as shown in Figure 12 for a node having four states. Belief update (eq. (15)) and prior support to child nodes (eq. (19) are similar since they use multiplication operations as well. Prior estimation operation represented by eq. (17) can be implemented using add-multiply composers, as shown in Figure 13. Likelihood support to parent node (eq.

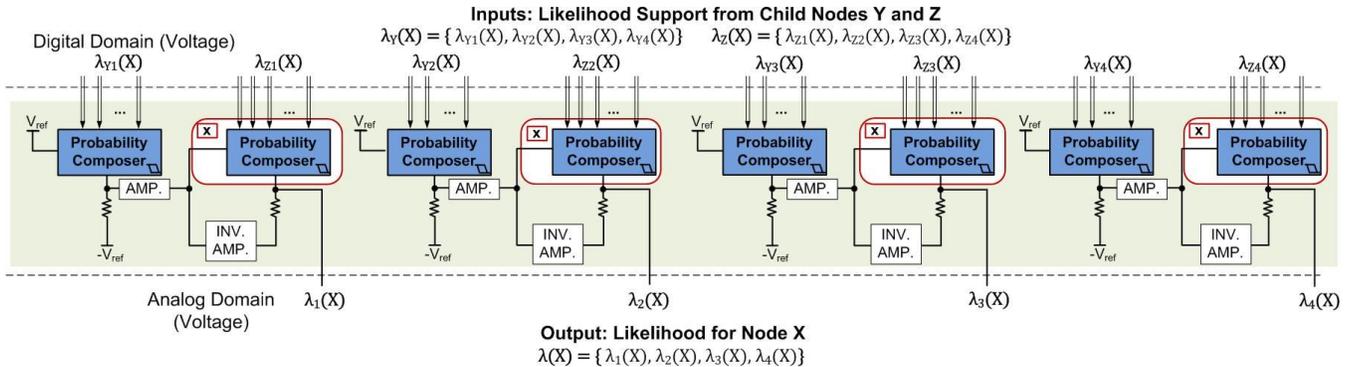

Figure 12. Implementation *of likelihood estimation* operation using Multiplication Composers, corresponding to eq. (16). The inputs to the module are likelihood support message vectors **λ_Y(X)** and **λ_Z(X)** from child nodes Y and Z respectively. Each vector has four elements corresponding to each state of node X.

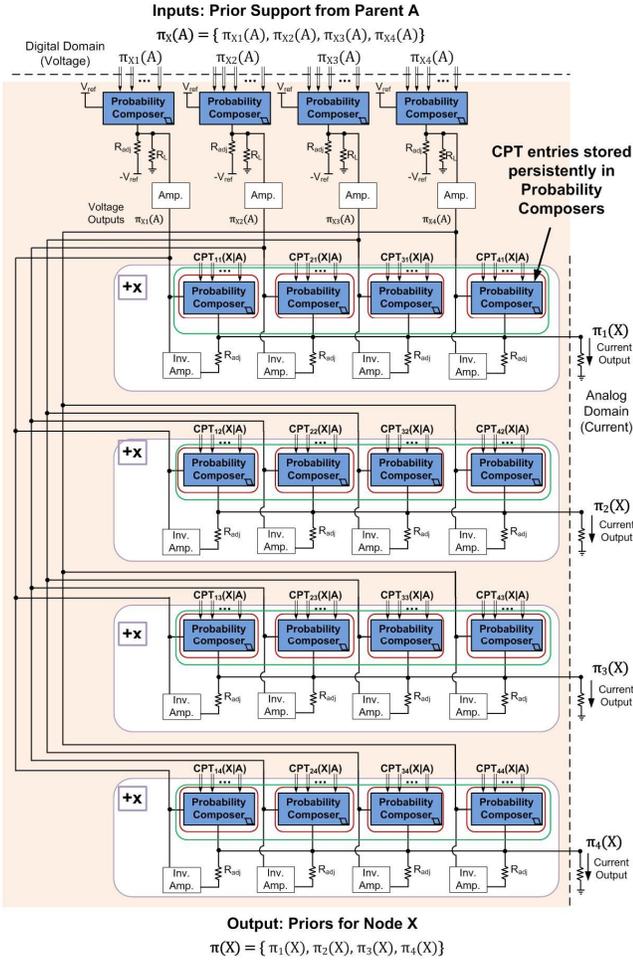

Figure 13. Implementation of *prior estimation* operation using Add-Multiply Composers, corresponding to eq. (17). The inputs to the module is a prior support message vector $\boldsymbol{\pi_X(A)}$ from parent node A. Each vector has four elements corresponding to each state of node X. The conditional probability table (CPT) elements are stored persistently in the Probability Composers as indicated, obviating the need to interface with external memory.

(18)) is implemented similarly.

VII. METHODOLOGY AND COMPARISON WITH BOOLEAN LOGIC APPROACH

HSPICE circuit simulator was used to verify the operation of Composer circuits with binary S-MTJ devices. The S-MTJ device characteristics (Figure 2) were used to build HSPICE behavioural macromodels, using voltage controlled resistors incorporating the data points, and custom voltage controlled delay elements for modeling switching delays. For the non-volatile device, a digital flip-flop was used to encode the S-MTJ resistance state persistently.

Elementary Arithmetic Probability Composers were implemented and functionally verified with HSPICE. Probability Composers with 10 S-MTJs were used in each Elementary Arithmetic Composer topology to enable an output resolution of 0.1, inspite of using binary S-MTJs. Analog CMOS support circuits (e.g. op-amps) were behaviorally modeled for the correction circuits and voltage adjusters. The S-MTJ device may be engineered to have a high $R_{OFF}/R_{ON}$ such that ε-factor can be neglected, obviating the need for correction circuits.

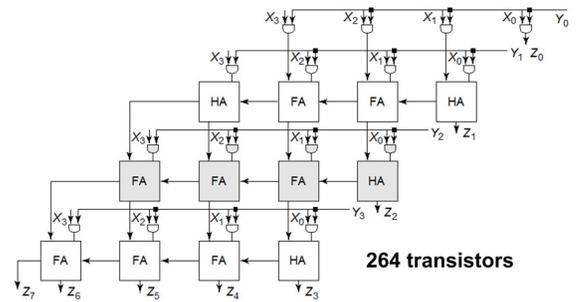

Figure 14. A 4-bit Boolean multiplier with a resolution of 1/8.

For evaluation and comparison with Boolean logic approach, we used an example of implementing Likelihood Estimation operation for Bayesian Inference. Here, every variable was assumed to support a maximum of 4 states, since target applications such as gene expression networks typically require three states for describing discrete gene expression levels [19]. This operation is described by eq. (16) and involves four multiplications to be performed between likelihood support messages from child nodes (probabilities). Likelihood estimation operation was implemented with four Multiplication Composers, each having a resolution of 1/10 (see Figure 12). Area was calculated based on the S-MTJ device dimensions and the spacing required between them to minimize magnetic interactions. Detailed simulations revealed a center-center spacing of 500nm for the S-MTJ devices. Using this, the rectangular area of each S-MTJ device with spacing was calculated to be $0.25\mu m^2$. Area for CMOS analog support circuits was estimated based on area requirement of a wide-range differential transconductance amplifier [18] in 45nm technology. Latency and average active power for Composer circuits were extracted using HSPICE simulations. The CMOS analog support circuits were behaviorally modeled in HSPICE, and worst case delay was estimated to be around 100ns for a CMOS amplifier (calculated based on HSPICE simulation of a wide-range differential transconductance amplifier [18] driving a load of 10 S-MTJs). Similarly, average active power was extracted for the transconductance amplifier circuit and used to estimate the power dissipation of analog support circuits.

We implemented the Likelihood Estimation operation using 4-bit and 5-bit CMOS Boolean multipliers, providing resolution of 1/8 and 1/16 respectively for comparison. Figure 14 shows a conventional Boolean implementation for an array-based multiplier using 4-bits (for a resolution of 1/8). The CMOS multipliers were described using RTL-level Verilog HDL and gate-level designs were synthesized with Synopsys Design Compiler. The physical layout design was extracted from the gate-level design using Cadence SoC Encounter for determining physical area. The HSPICE netlists for the multipliers, including parasitic resistances and capacitances due to routing, were generated using Cadence Virtuoso and 45nm North Carolina State University (NCSU) Product Development Kit (PDK) library. Performance and average

TABLE II. COMPARISON OF BOOLEAN VS. PROBABILITY ARITHMETIC COMPOSER FRAMEWORK FOR IMPLEMENTING LIKELIHOOD ESTIMATION

| Likelihood Estimation Operation for Bayesian Inference | | Area (µm²) | Active Power (mW) | Latency (µs) Computation | Latency (µs) Memory Access |
|---|---|---|---|---|---|
| *45nm CMOS Boolean: Four array-based multiplier modules* | *4-bit Multipliers w/ Resolution 1/8* | 1920 | 2.92 | 0.0005 | 10 |
| | *5-bit Multipliers w/ Resolution 1/16* | 3080 | 4.4 | 0.00065 | 10 |
| *Probability Arithmetic Composer: Four Multiplication Composers* | *Composers w/ Resolution 1/10* | 24.32 | 0.016 | 0.144 | NA |

Note: Memory access latency estimated based on state-of-the-art flash solid state drive read latency.

active power for CMOS multipliers were calculated using HSPICE simulations.

The Probability Arithmetic Composer implementation showed 79x area benefit compared to 4-bit CMOS multipliers, and 127x area benefit compared to 5-bit CMOS multipliers. This is due to compact mixed-signal implementation of the underlying operations, inspite of the large spacing between devices for minimizing magnetic interaction of free layers. Shielding techniques [20] may be used to further increase the density of S-MTJ based Composer circuits. In terms of active power, the Probability Arithmetic Composers showed 142x lower active power vs. 4-bit CMOS multipliers and 214x lower power compared to 5-bit CMOS multipliers, due to the use of far lesser active devices. Performance of Probability Arithmetic Composer circuits for computation showed 288x degradation (higher delay) compared to 4-bit and 221.5x degradation compared to 5-bit CMOS multipliers. The performance of Composers in current designs was limited mainly by the high parasitic capacitance associated with S-MTJ devices (in the order of 1fF). The performance may be improved through further research in the choice of materials used for reducing the parasitic capacitance of these devices.

However, it is important to note that when considering memory access overhead for CMOS using non-volatile flash storage (typically having read access latencies of the order of 10µs even with state-of-the-art solid-state drives), the overall performance of Arithmetic Composer approach improves by ~70x over CMOS due to its memory-in-computation feature. While some of the CMOS performance degradation may be mitigated by using SRAM caches, this approach would face increasing complexity and memory bottleneck issues due to limited bandwidth and resource contention, when scaling to large problems involving variables in the order of a hundred thousand to a million (which is possible when considering genetic interactions and gene-environmental interactions for example). Composer circuits, on the other hand, can easily scale to large number of variables through memory-in-computation since they incorporate non-volatile storage in S-MTJ resistance states, thus obviating the need to interface with an external memory.

Since Composer circuits are non-volatile, they can be completely switched-off after computation thereby eliminating static power dissipation. We envision that a large BN implemented using Probability Arithmetic Composer framework would operate asynchronously, such that at any given event step in a sequence of operations only the active nodes are operational. All other inactive nodes will be switched off to mitigate power dissipation. These benefits are expected to improve further through the development and use of multi-valued S-MTJ devices.

VIII. CONCLUSION

A novel non-Boolean magneto-electric nanocircuit technology framework was presented for nanoscale cognitive reasoning based on probabilistic graphical models, such as Bayesian Networks. HSPICE was used to simulate and verify the operation of the proposed Arithmetic Composer circuits, where arithmetic functions themselves were used as building blocks rather than Boolean logic. It showed up to 127x area benefit, 214x lower active power and 70x lower latency (when memory access overhead is taken into account) vs. CMOS Boolean implementation of likelihood estimation operation used in Bayesian Network inference. The benefits of the proposed approach are due to several factors: (i) Data representation is directly on probabilities. (ii) Novel magneto-electric Arithmetic Composer circuits allow direct arithmetic operations on probabilities without emulation through logic. (ii) Non-volatility in S-MTJ resistance states is leveraged for novel memory-in-computing schemes, eliminating the need for interfacing with external memory and transfering data between memory and computational units. From an architectural perspective, the integration of memory in computation potentially overcomes the memory bottleneck in conventional stored-program approach with CMOS. While S-MTJ devices were used in this paper, the Probability Arithmetic Composer paradigm is generic and may be used with other devices that exhibit multi-domain interactions with non-volatility [21].

Future work in this direction will look at implementing learning functions using the Probability Arithmetic Composer framework. Several architectures may be possible using these Composers; one direction is to implement a Bayesian memory incorporating Composers such that it helps in accelerating Bayesian inference and learning, or a distributed architecture directly mapping probabilistic graphical models in hardware can be envisioned that involves a parallel implementation of computations involved in inference and learning. It could lead to highly efficient cognitive reasoning machines at nanoscale in terms of area, power and performance, compared to conventional implementations used today.

DISCLAIMER